# Laser-induced fluorescence study of medieval frescoes by Giusto de' Menabuoi

Roberta Fantoni [a],[*], Luisa Caneve [a], Francesco Colao [a], Luca Fiorani [a], Antonio Palucci [a], Ramiro Dell'Erba [b], Vasco Fassina [c]

[a] ENEA Technical Unit for the development of applications of radiations, CR Frascati, V. E. Fermi 45, 00044 Frascati, Italy
[b] ENEA Technical Unit technologies for energy and industry – Robotics Laboratory, CR Casaccia, Via Anguillarese 301, 00123 Roma, Italy
[c] Soprintendenza ai Beni Storici Artistici ed Etnoantropologici per le province di Venezia, Belluno, Padova e Treviso, Rio Marin, 30124 Venezia, Italy

## a b s t r a c t

Laser-induced fluorescence (LIF) is a powerful remote and non-invasive analysis technique that has been successfully applied to the real-time diagnosis of historical artworks. Hyperspectral images collection on fresco's and their false colours processing allowed to reveal features invisible to the naked eye and to obtain specific information on pigments composition and consolidants utilization, the latter also related to former restorations. This report presents the results obtained by ENEA LIF scanning system during a field campaign conducted in June 2010 on fresco's by Giusto de' Menabuoi in the Padua Baptistery. The data collected by LidArt allowed the detection of Paraloid B72 and Movilith/Primal AC33, guiding the restorers in their conservation actions.

## 1. Introduction

Laser-induced fluorescence (LIF) is a powerful remote analysis tool that has been successfully applied to the real-time diagnosis of historical artworks, allowing the observation of features invisible to the naked eye, as pigment composition [1], biological attack [2] and trace of former restorations.

In a typical LIF instrument, an ultraviolet (UV) laser beam irradiates a sample and an optical system measures the fluorescence spectrum that contains information on the target composition. Nevertheless, alternative spectroscopic techniques, such as XRF and micro-Raman are suitable to obtain analogous in situ information [3]. LIF is up to now the only one for which remote application has been demonstrated up to ten of meters. LIF is indeed fast, remote [4], non-invasive, sensitive and selective. These advantages have encouraged its application in the real-time monitoring of historical frescoes, mosaics, paintings and stones. Former studies showed its high potential as a diagnostic tool for cultural heritage either operating in spectral [1,2] or time [5] resolved mode.

The Technical Unit for the development of applications of radiations, part of the Italian National Agency for New Technologies, Energy and Sustainable Economic Development (ENEA) has worked in this field since 2003 and has developed various prototypes. The system here described (LidArt) is a lidar fluorosensor (optical radar based on LIF) and its first version was patented in 2007 [6]. At present, it is even more compact (contained in a cylinder of radius 29 cm and height 18 cm), hyperspectral and time resolved (the fluorescence spectrum is measured with a wavelength resolution of few nm and a time resolution of few ns). Moreover, it is light, robust, transportable, user-friendly and cost-effective. Eventually, its scanning (the artwork surface is probed remotely aiming the laser beam) is based on an innovative approach [7] that reduces the acquisition time from hours to minutes.

Thanks to sophisticated data processing techniques as false-color imaging, principal component analysis (PCA) on spectra and spectral angle mapping (SAM) on images, LidArt has detected characteristics invisible to the naked eye, as pigment compositions (e.g. titanium white vs. zinc white [8]), pigment diffusions (lime and casein) [9], biological attacks (algae and fungi) [10], consolidants (usually resins) [11], deteriorations, depigmentations, retouches and varnishes.

In particular, restorers need a precise knowledge of surface materials once engaged in removing traces of former restorations. LIF images, collected in the Padua Baptistery and processed in order to retrieve information on constituent materials, are here presented and discussed. The field campaign was carried on in June 2010 prior to the planning of successive restorations. To this aim, the investigation was focused onto the recognition of consolidants and binders, with the support of an available data base and the development of statistical tools for main spectral bands identification.

The construction of Padua Baptistery, located on the right side of the cathedral, started in 12th century and was accomplished in 1281. Its fresco's are considered Giusto de' Menabuoi's masterpiece. In the present study, carried on within a contract signed by ENEA and *Soprintendenza ai Beni Storici Artistici ed Etnoantropologici per le province di Venezia, Belluno, Padova e Treviso*, selected areas of the Paradise fresco on the central dome and of the Genesis on the tambour have been investigated by means of the ENEA LIF scanning system.

[*] Corresponding author.
E-mail address: fantoni@frascati.enea.it (R. Fantoni).

After a short description of the last LidArt prototype, results obtained during the field campaign conducted from June 7 to 11 are presented and discussed. Some conclusions on the appropriateness of the developed technological tool to the required remote investigation on a painted CH surface, are given at the end.

## 2. Experimental

### 2.1. The set-up

The LIF scanning system developed and patented [6] at the ENEA laboratory in Frascati is capable to collect hyperspectral fluorescence images scanning large areas for applications to cultural heritage surfaces (e.g. frescoes, decorated facades etc.). A new compact set-up has been built, and successively patented [7], aimed to increase the performances in terms of space resolution, time resolved capabilities and data acquisition speed. Major achievements have been reached by a critical review of the optical design and consequently of the detector utilized: the point focalization mode has been changed into a line focalization by using a quartz cylindrical lens and a imaging spectrograph (Jobin-Yvon CP240); the linear array detector, responsible for the multichannel spectral resolution, has been replaced with a square ICCD sensor (ANDOR iStar DH734, pixel size 13 μm), mounted behind a slit parallel to the laser line footprint during the scanning.

This arrangement is characterized by having the spatial and spectral information on two mutually orthogonal directions imaged on the detector, with submillimetric spatial resolution and a spectral resolution better than 2 nm. Additionally, it is possible to implement time resolved measurements on the nanosecond scale by gating the spectral detector.

The overall current system performances are horizontal resolution 640 pixel, 0.1 mrad angular resolution, minimum acquisition time per line 200 ms, field of view (FOV) aperture 5.7° (corresponding to a scanned line of 2.5 m at 25 m distance). With the latter optics, an image of $1.5 \times 5\,m^2$ is scanned in less than 2 minutes at 25 m.

The compact arrangement of the experimental apparatus is shown in Fig. 1, together with the adopted optical scheme. The Nd:YAG compact laser source could be operated at two different wavelengths in the UV: 355 nm or 266 nm, 1 mJ of output energy were produced in short pulses (10 ns) at 20 Hz. Special care has been paid to check on reference samples in laboratory that the utilized laser fluence was low enough to avoid any photo-damage induced by the laser to the painted surface (energy density < 7 μJ/cm$^2$) as tested on fresco specimens to the purpose realized. In particular measurements were carried defocussing the laser beam. The combination of (low) laser repetition rate with (high) scanning speed allowed to avoid the occurrence of any photochemical process at the surface (e.g. pigment oxidation or consolidant decomposition) related with continuous irradiation of the same points.

### 2.2. Operating modes

The line scanning system can operate in two operating modes by slightly changing the active components in the entire device: i.e. reflectance and fluorescence measurements. These modes share indeed most of the optoelectronic equipment, namely the collection optics and the spectral detector, while differing, on one hand, in the way the sample is excited and, on the other hand, in the way the spectral detector is operated.

Reflectance measurements are performed in a scan in which the laser is switched off while the sample is exposed to the light emitted by a NIST traceable lamp. The collection optics are focussed onto the scanned surface and the acquisition camera is running by locking its internal synchronism to the power line (to avoid image flickering). The result gives the reflectance spectrum for each pixel of the scanned area, from which the CIE/lab coordinates can be computed once the system is calibrated against a reference surface.

The acquisition of fluorescence spectra induced by laser is carried on during a scan with the laser switched on. The system uses a procedure to record spectra also in full day light: in this case, a proprietary method is used to discriminate between the light induced from the laser (LIF signal) with respect to the light diffused by the area under study and by the environment. The result gives the fluorescence spectrum for each pixel of the scanned area; then a successive data analysis of the acquired spectrum provides detailed information on the area under study.

LIF fluorescence measurements require the adoption of a peculiar measurement protocol in order to achieve a good detecting efficiency and high spatial resolution even in the very broad spectral range of radiation emitted upon laser excitation at 266 nm. In fact, because of the use of refractive collecting optics, the focal planes for ultraviolet (UV) and visible (VIS) radiation occur in different position, causing defocus effects and degrading the spatial resolution in either of the regions. To overcome the problem, successive separate acquisition runs are scheduled for every scan, each of them optimized for a specific wavelength range.

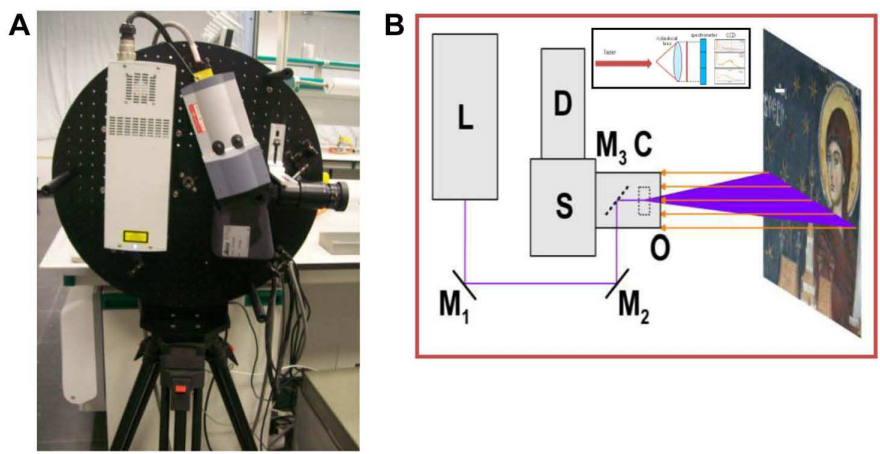

**Fig. 1.** Compact LIF line scanning: (A) picture, right scheme (B), the vertical optical bench is mounted on a tripod, vertical scanning is performed by means of an accurate stepping motor on the back side. Transmitting and receiving optics are in the front (M1, M2, M3, C, O), the CCD (D) with the spectrograph (S) and the laser (L) are on the rear.

Actually, three consecutive acquisition runs are scheduled: in the first the collecting optics is focused on the UV region from 250 to 450 nm; the successive two scans are focused on the spectral region from 400 to 750 nm to acquire respectively the fluorescence induced in the visible and the reflectance image.

### 2.3. Analysis of spectral measurements

To speed up the processing of acquired images, the most relevant spectral features are identified by Principal Component (PC) analysis. Although it is commonly admitted that the PCs do not possess any direct physical meaning, they can nevertheless be represented as spectra suitable to be described in terms of bands. In some cases, a given PC corresponds to well defined spectroscopic bands with associated peaks, while in other cases more complicated trends and shapes are observed: most frequent is the case of a band set against another. Few of the PCs are usually retained for subsequent analysis: typically five to eight components are enough to describe the entire spectral data set. It is also worth noticing the possibility to build suitable linear combinations of the computed PCs to have a faithful representation of each pixel spectra, eventually used for the computation of standard CIE/lab colorimetric measurements.

In the present paper, the PC analysis is devoted to the identification of prominent spectral features, thus relieving from the lengthy time consuming examination of each acquired spectrum [12]. This procedure is advantageous because it is fast and can run in a semi-automatic mode. However, it has the inconvenience to give a global analysis, possibly ignoring those local peculiarities which do not possess enough statistical significance. To overcome this drawback local PC analysis are also performed on different portion of the scanned areas, then the results are analyzed separately. Once identified, spectral bands are sought for in the acquired LIF spectra, completing the data analysis.

A different method, used in the analysis of spectral images, concerns the identification of regions having a specific spectral content. Typical is the case of identification of a given pigment in an image: such task is accomplished either by a band analysis, or by using projection algorithms like Spectral Angle Mapper (SAM) or Spectral Correlation Mapper (SCM) in combination with an available data base [13]. Although the mapper algorithms perform well with a low computational cost, their performances are generally lower with respect to the band analysis procedures.

## 3. Results: LIF reference spectra from fresco materials

Before presenting images remotely collected on frescoes at the Baptistery, in order to proceed with an effective data analysis, it is worthwhile to report the reference spectra which can be utilized for SAM projections in order to identify specific contribution for different chemicals at the fresco surface. To this aim laboratory samples were characterized as prepared according to traditional receipts concerning both materials and methods. A final layer of consolidant has been added on top of each sample surface. Based on former information of restorations carried on in Padua Baptistery in the following, we describe only a part of the available data base, including acrylic resins, binders and pigments, which are most likely relevant to the frescoes upon investigation. Note that normalized spectra, gather after different sample preparation resulted always very well comparable, thus supporting the peculiar spectral feature associated to each component upon monochromatic UV excitation.

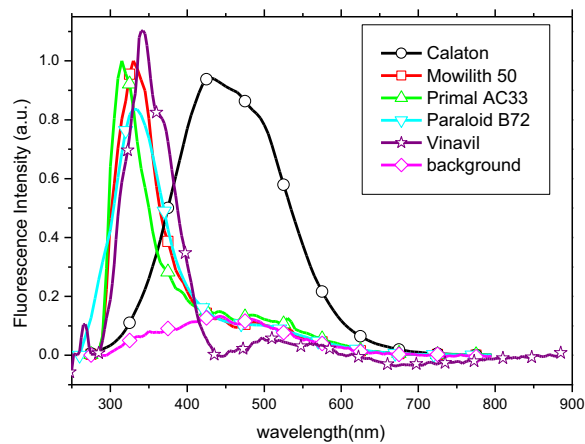

**Fig. 2.** Fluorescence spectra of acrylic resins on laboratory frescos samples excited at 266 nm.

### 3.1. LIF reference spectra of acrylic resins

Since about 40 years acrylic resins were commonly utilized as consolidant in restoration of painted surfaces, including frescoes. LIF spectroscopy has been utilized for the characterization of five different resins: paraloid B72, calaton, primal AC33, mowilith 50 and vinavil. A laboratory sample set was prepared during a cooperation Ljubljana Restoration Centre. Each sample was formed by a plaster substrate, a pigment layer and covered by an acrylic resin at different thickness. Reference spectra were collected upon excitation at 355 nm and 266 nm, the former showed a broad blue-green band characteristic for the entire family, while the spectra collected from the latter are shown in Fig. 2, their main features are summarized in Table 1.

The spectra collected after excitation at 266 nm are suitable to resins classification, since they show specific emission bands for each different sample [11]; namely only calaton presents a characteristic blue-green broad visible emission, while all others are strong UV emitters. Among them vinavil can be recognized by its slightly broader UV band accompanied by a weaker well defined second peak at 510 nm. Finally, the substrate emission alone is completely different, not showing any signature overlapping with the examined consolidants.

### 3.2. LIF reference spectra of binders

Since historical times, organic binders are utilized to keep together pigments powders. Eight different compounds, listed in Table 2, were interrogated by LIF at two different excitation wavelengths. Samples were realized at the Chemistry Dep. of Rome University Sapienza, by placing directly the binder on top of a non fluorescent substrate.

Reference spectra, collected at 2 m distance upon laser excitation at 355 nm, are shown in Fig. 3, respective band centre positions are given in Table 3. Data obtained upon excitation at 266 nm are not shown since, a part from rabbit glue peaked at 310 nm, all spectra appear very similar with a maximum around 340 nm and a tail along the visible. We may thus conclude that, once information on binders are searched for, it is worthwhile to utilize the longer UV wavelength (355 nm) to detect spectral features which might allow for their identification.

**Table 1**
Band centres of fluorescence spectra of acrylic resins on laboratory frescos samples excited at 266 nm.

| Resin | Band centre (nm) (exc) = 266 nm |
|---|---|
| Calaton | 435 |
| Mowilith 50 | 330 |
| Primal AC33 | 315 |
| Paraloid B72 | 335 |
| Vynil | 340 |

**Table 2**
Pigments peak emission excited at 266 nm.

| Pigment | Band centres (nm) (exc) = 255 nm |
| --- | --- |
| Line oil | 460 |
| Lead white | 460 |
| Calcite | 450, 580 |
| Lead sulfate | 450 |
| No pigment (plaster) | 460 |

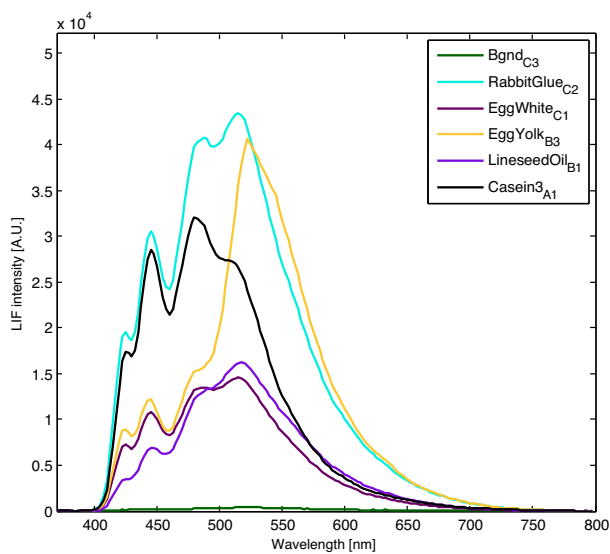

**Fig. 3.** Fluorescence spectra of binder on an inert substrate upon excitation at 355 nm.

## 4. Results: the campaign on Padua Baptistery frescoes

### 4.1. The scanned areas

In the Baptistery, different portions of the frescoes have been scanned within the dome and on the tambour. Six images were collected on the dome vault and eight images on the tambour. Data were acquired at 15 m distance from the dome at a slightly smaller distance from the tambour. Images collected are shaped as rectangular stripes about 1.5 m broad and typically from 8 m to 12 m long.

**Table 3**
Binders peak emission excited at 355 nm.

| Binder | Band centres (nm) (exc) = 355 nm |
| --- | --- |
| Casein | 450 |
| Line oil | 520 |
| Egg white | 520 |
| Rabbit glue | 520 |
| Red white | 540 |
| Background (silica) | 520 |

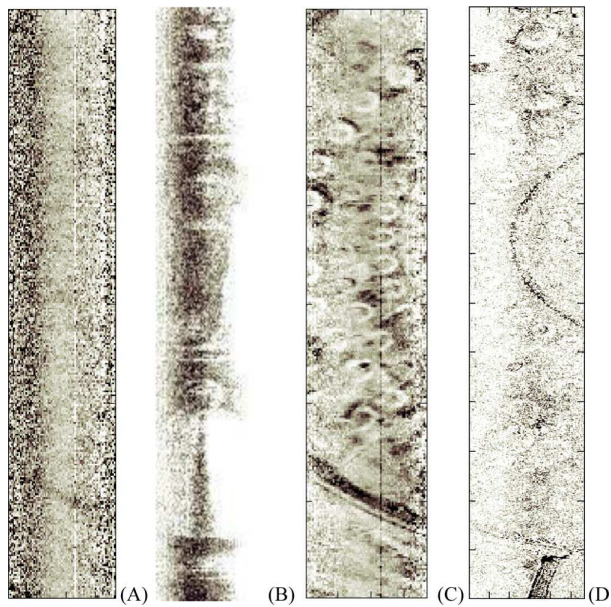

**Fig. 4.** Sections of the dome vault scanned by LIF, grey levels correspond to different emission intensity (black represent the highest intensity); (A) and (B) band at 293 nm; (C) and (D) band at 370 nm.

Data were acquired operating the set-up both in reflectance and fluorescence acquisition mode, the latter upon excitation at 266 nm. Data were processed by means of an automatic statistical analysis (PCA) aimed to recognize the presence of prominent bands. Once major fluorescence bands have been identified, their assignment has been checked against the available consolidant data base by means of SAM projections, and the relevant distribution on the image has been obtained.

The band analysis, upon which the monochromatic image reconstruction reported in Figs. 4 and 5 is based, derives directly from the PCA results. In particular the narrow UV bands at 293 and 370 nm have been considered in Fig. 4. On the left (A and B insets), the absolute intensity measured at 293 nm is shown, which appear to generate a quite uniform distribution on the entire images. Upon the assumption of a linear dependence coupling the quantitative distribution of fluorophores to the respective peak intensity, we infer a corresponding uniform distribution of the compound. A similar situation has been observed for all scanned areas in the dome (images not reported), thus supporting a diffuse and homogeneous use of the compound characterized by the emission at 293 nm, in general compatible with acrylic resins (e.g. primal AC33) [11,14]. On the right side of Fig. 4 (C and D insets), the absolute intensity measured at 370 nm is shown, which is peaked on well-localized areas. Concerning the identification of the compound to which this emission could be ascribed, although several compounds do present this signature, most probable candidates among consolidants are vinyl compounds (e.g. mowilith) which emit at longer wavelength with respect to acrylic resins [11,14]. Nevertheless, data in Fig. 6 allow for both the precise localization and the semi quantitative determinations of materials present as coating on the examined surface, their univocal assignment to a specific chemical compound is not possible on the basis of remote LIF spectra alone; the latter would require additional local measurements with complementary techniques (e.g. micro-stratigraphy, XRF, micro-XRD [15], etc.) to be carried on one or a few specific points of the investigated areas.

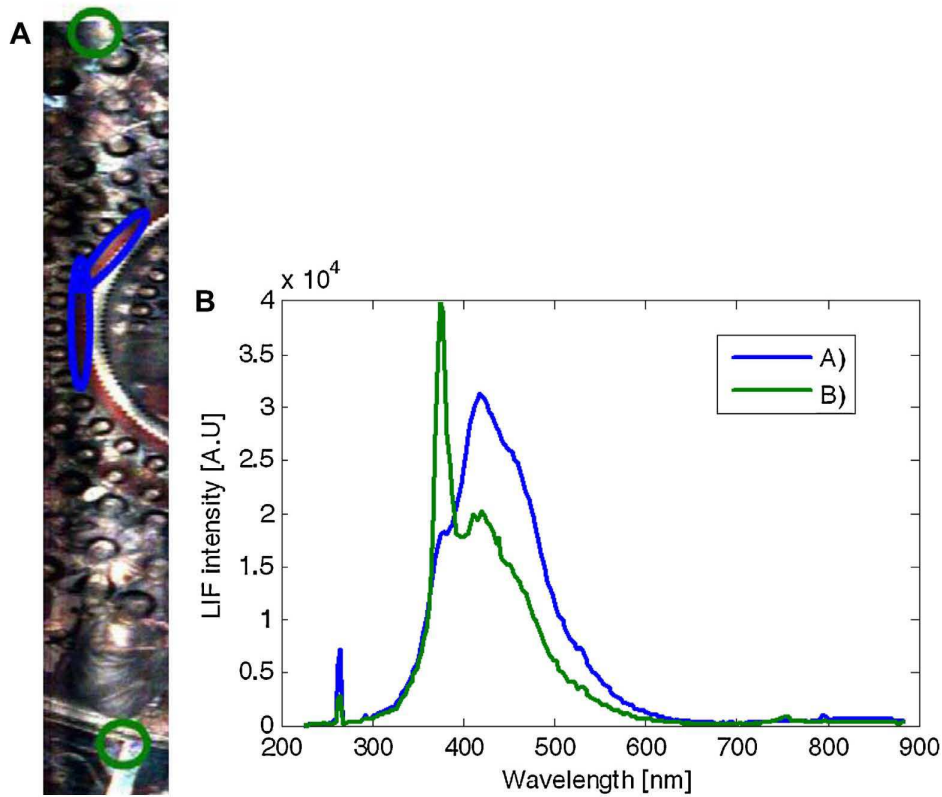

**Fig. 5.** (A) False colour RGB reconstruction of a section of the dome vault based on bands at 293 nm, 370 nm and 550 nm; (B) average spectral content in selected spectral regions, which are marked on the left image. Note the intense band appears at 375 nm tentatively assigned to vinyl consolidants.

A further qualitative results of LIF scanning can be obtained combining three significant bands in a single false colour images, an example of this elaboration is presented in Fig. 5 (for the monochromatic image shown in Fig. 4D). The inset (A) show the RGB reconstruction based on band peaked at 550 nm, 370 nm and 293 nm; the inset (B) shows the average spectrum in the respective circled area on the left, significant emissions at 375 nm and at 293 nm (confirming PCA results) are observed, their possible assignment to different consolidants being above discussed.

The SAM application gives a precise localization of different materials, however this is not true when overlapping contributions are found with comparable intensities. An example of this unfavourable case is shown in Fig. 6: the left inset (A) shows the similarity map referred to paraloid B72 fluorescence spectrum, covering about 15% of the examined surface, however an analogous percentage is obtained when considering the substratum spectrum in the right inset (B). The remarkable space correlation among the retrieved map is an additional indication of the strong overlap between the considered spectral contribution which cannot be effectively disentangled by the used SAM algorithm.

As a last example of consolidant identification, a vinyl compound, whose fluorescence spectrum is available as pure compound (Fig. 2), has been considered. Since the compound could not be deposited on a plaster similar to the one utilized in Padua Baptistery and it was necessary to have a good spectral reference to retrieve a reliable space distribution, a two-step analysis was carried on. First, the similarity map was produced with respect to the pure compound, small areas (about 1%) unambiguously characterized by its presence were identified (Fig. 7A). Second, the spectral signature of these areas was assumed as internal reference of the compound on the used plaster and the final distribution was obtained (Fig. 7B), covering about 11% of the examined surface. Although this indirect procedure might give rise to artefacts, in case of initial incorrect assumption or noisy spectra, and the confirmation of the assignment should be achieved by means of complementary techniques whenever possible, the proposed method is a valid tool for an immediate non-invasive screening and also to the most effective localization of successive sampling areas.

### 4.2. Reflectance measurements on pigments

An RGB image obtained from reflectance measurements on the vault is shown in Fig. 8A. Spectral features have been identified by PCA The reconstructed reflectance image reveals an interesting feature under Christ's right hand: a red area with highly saturated color, whose spectral analysis shows an alteration in the ratio between the red and blue maxima (at 622 nm and 477 nm, respectively).

The complete space location of the spectral anomaly has been recognized by means of SAM analysis based on an internal reference end-point in the area of maximum contrast. The result of SAM analysis is reported in Fig. 8, for the red peak (B), for the blue band (D) and for the anomalous bimodal distribution (C) observed on the red painted area. In particular, the F450 band analysis does not define sharp border for pigmented areas (see Fig. 8) because of an interfering peak at 477 nm appearing both in blue and red pigment spectra (Fig. 8); on the other hand, the peak ratio F620/F477 allows to precisely localize features with spectral differences of the order of at least 5%. The physical meaning of the observed anomaly is not clear, since different causes might contribute to its occurrence: retouches with a different red pigment, humidity on the wall, different binders or added varnishes with diffusive inhomogeneous reflectance at the surface. Once again, additional complementary measurements would be necessary for a definite answer, however, the present measurements indicate where samples should be collected and how much the anomaly extends on the painted surface.

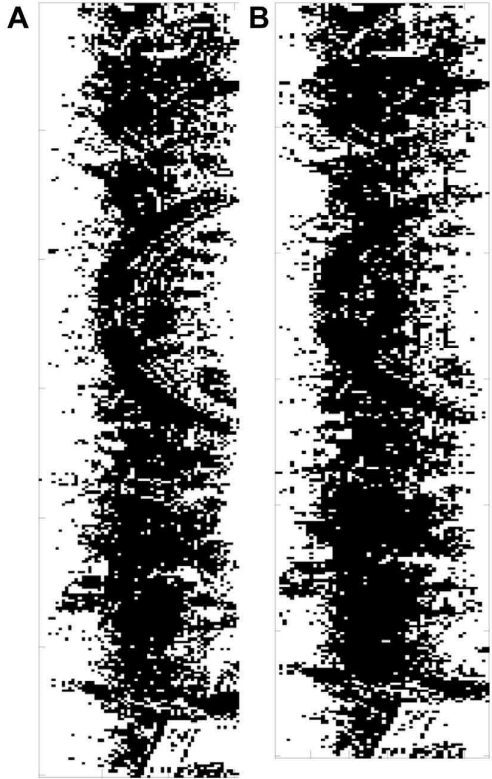

Fig. 6. SAM maps associated with: paraloid B72 (A) and plaster (B).

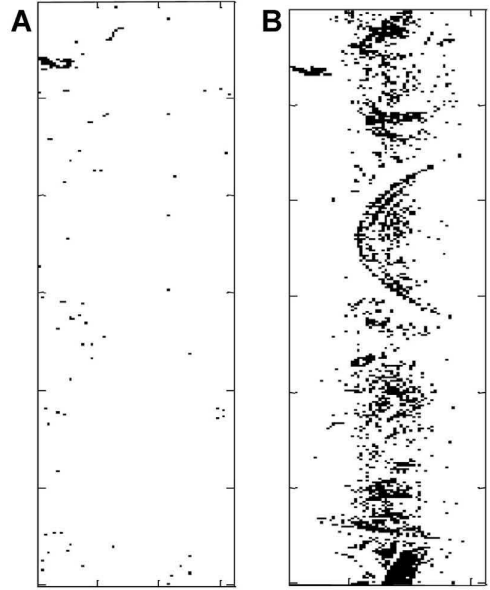

Fig. 7. SAM maps associated to: vinyl from our data base (A) and a vinyl with an internal reference (B).

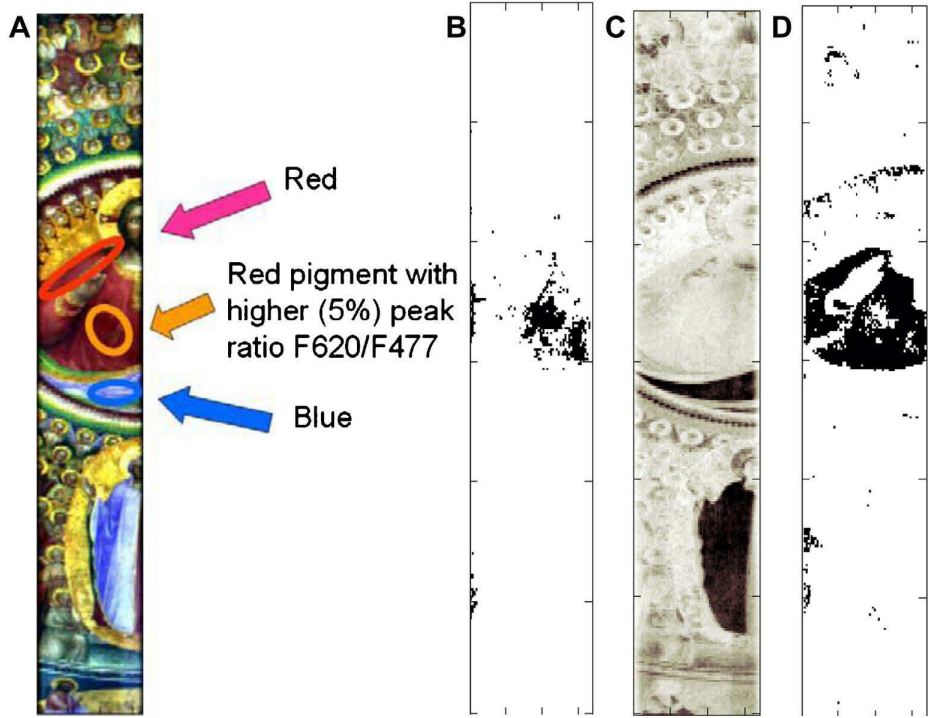

Fig. 8. (A) RGB reconstruction of reflectance image collected on the dome; SAM results for (B) red peak; (C) bimodal spectrum; (D) single band at F450 nm.

## 5. Conclusions

Laboratory measurements, aimed to the realization of a LIF data base for remote investigation of frescoes, in combination with the developed image analysis methodology, have been utilized to discuss results collected during the campaign of characterization of Padua Baptistery frescoes, as an example of monumental painted surface. Present results allow to summarize major achievements of the proposed innovative technique, as follows:

- the LIF instrument developed is suitable to non-invasive and remote surface diagnostics; namely it permits to scan large areas with medium/high resolution and to release digital images whose false colors can be associated with detected spectral features observed upon UV excitation at 266 nm (e.g. presence of specific consolidant/binder);
- the identification of the consolidant/binder, is possible provided that its LIF spectrum is present in the available data base as detected on a similar complex matrix; alternatively the presence of a class of consolidants (e.g. acrylic, vinyl, etc.) can be assumed and its distribution can be measured on the entire image by using as reference and internal standard whose spectrum has been obtained in an area of the image where it dominates;
- the additional availability of reflectance images, measured upon external illumination or utilizing an additional visible lamp while keeping the laser switched off, by means of the same scanning system, gives information on pigments distribution and degradation.

The limitations of the diagnostic technique rely mostly on the nature of the CH target considered, due to its complexity with a multilayered structure often containing materials with small chemical differences. Major drawbacks for frescoes are reported in the following list:

- difficulties to built a reliable data base, containing all original materials prepared according to specific historical receipts and on documented restorations (the latter for consolidants added in modern times);
- modest differences in spectral signatures, either related to chemical similarity, or to broadband emission characteristic of fluorescence for solid surfaces, strongly altered by the fluorescent substrate (e.g. plaster), which would prevent to get improved results by hardware implementation of high spectral resolution acquisition.

Due to the mentioned limitations, a quantitative LIF scanning remote analysis seems not to be feasible also in perspective. Conversely, the advantages offered by fast and semi-automatic analysis capable to recognize anomalies in the distribution of surface components (either consolidants, binders, or even some pigments) are of interest to the conservator who is aware where to proceed with accurate point analyses utilizing either different spectroscopic (Raman, LIBS), Spectrometric (SIMS), or imaging (IR, XRF) techniques to have definite material identification and respective quantitative analysis, the latter suitable for renormalization of LIF scanning intensity distribution on remotely collected images.

## References

[1] D. Anglos, M. Solomidou, I. Zergioti, V. Zaffiropulos, T.G. Papazoglou, C. Fotakis, Laser-induced fluorescence in artwork diagnostics: an application in pigment analysis, Appl. Spectrosc. 50 (1996) 1331–1334.
[2] D. Lognoli, G. Lamenti, L. Pantani, D. Tirelli, P. Tiano, L. Tomaselli, Detection and characterization of biodeteriogens on stone cultural heritages by fluorescence lidar, Appl. Opt. 41 (2002) 1780–1787.
[3] J. Striová, G. Coccolini, S. Micheli, C. Lofrumento, M. Galeotti, A. Cagnini, E. Castellucci, Non-destructive and non-invasive analyses shed light on the realization technique of ancient polychrome prints, Spectrochim. Acta A 73 (2009) 539–545.
[4] G. Cecchi, L. Pantani, V. Raimondi, L. Tomaselli, G. Lamenti, P. Tiano, R. Chiari, Fluorescence lidar technique for the remote sensing of stone monuments, J. Cult. Herit. 1 (2000) 29–36.
[5] D. Comelli, C. D'Andrea, G. Valentini, R. Cubeddu, C. Colombo, L. Toniolo, Fluorescence lifetime imaging and spectroscopy as tools for nondestructive analysis of works of art, Appl. Opt. 43 (2004) 2175–2183.
[6] F. Colao, R. Fantoni, L. Fiorani, A. Palucci, Fluorescence-based portable device based for the spatial scan of surfaces, in particular in the area of Cultural Heritage, Italian patent RM2007A000278 deposited on 21.05.2007.
[7] L. Caneve, F. Colao, L. Fiorani, A. Palucci, Portable laser radar system for remote surface diagnostics, Italian patent RM2010A000606 deposited on 17.11.2010.
[8] I. Borgia, R. Fantoni, C. Flamini, T. Di Palma, A. Giardini Guidoni, A. Mele, Luminescence from pigments and resins for oil paintings induced by laser excitation, Appl. Surf. Sc. 127 (1998) 95–100.
[9] F. Colao, R. Fantoni, L. Fiorani, A. Palucci, Application of a scanning hyperspectral lidar fluorosensor to fresco diagnostics during the CULTURE 2000 campaign in Bucovina, Revista Monumentelor Istorice/Review of Historical Monuments n. LXXV/1-2 (Bucharest, 2006) 53–61.
[10] F. Colao, L. Caneve, A. Palucci, R. Fantoni, L. Fiorani, Scanning hyperspectral lidar fluorosensor for fresco diagnostics in laboratory and field campaigns, in: J. Ruiz, R. Radvan, M. Oujja, M. Castillejo, P. Moreno (Eds.), Lasers in the Conservation of Artworks (LACONA VII), CRC Press, 2008, pp. 149–155.
[11] L. Caneve, F. Colao, R. Fantoni, L. Fornarini, Laser induced fluorescence analysis of acrylic resins used in conservation of cultural heritage. Proceedings of OSAV'2008, The 2nd Int. Topical Meeting on Optical Sensing and Artificial Vision, St. Petersburg, Russia, 2008, pp. 57–63.
[12] A.C. Rencher, Methods of Multivariate Analysis, Wiley Interscience, New York, 2002.
[13] G. Girouard, A. Bannari, A. El Harti and A. Desrochers, Validated Spectral Angle Mapper Algorithm for Geological Mapping: Comparative Study between Quickbird and Landsat-TM, Presented at the XXth ISPRS Congress, Geo-Imagery Bridging Continents, Istanbul, Turkey, 12–23 July 2004.
[14] C. Miliani, M. Ombelli, A. Morresi, A. Romani, Spectroscopic study of acrylic resins in solid matrices, Surf. Coatings Technol. 151–152 (2002) 276–280.
[15] S. Valadas, A. Candeias, J. Mirao, D. Tavares, J. Coroado, R. Simon, A.S. Silva, M. Gil, A. Guilherme, M.L. Carvalho, Study of mural paintings using in situ XRF, Confocal Synchrotron-mu-XRF, mu-XRD, Optical Microscopy, and SEM-EDS. The case of the Frescoes from Misericordia Church of Odemira, Microsc. Microanal. 17 (2011) 702–709.